\documentclass[fullpage]{article}

\newcommand{\product}[2]{ {#1} {\times} {#2} }

\newcommand{\triple}[3]{\mbox{$ \langle #1,#2,#3 \rangle $}}

\begin{document}

\title{\Large\bf Creating a Web Analysis and 
Visualization Environment}
\author{
  \renewcommand{\baselinestretch}{0.5}
  \small Robert E. Kent \\
  \small University of Arkansas \\
  \small Little Rock, Arkansas 72204, USA\\
  \small rekent@logos.ualr.edu\\  
  \and
  \small Christian Neuss \\
  \small Fraunhofer Institute for Computer\\ \small Graphics,
  \small Darmstadt, Germany \\
  \small neuss@igd.fhg.de 
}

	\date{}
	\maketitle
\begin{abstract}
Due to the rapid growth of the World Wide Web, resource discovery
becomes an increasing problem. As an answer to the demand for information management,
a third generation of World-Wide Web tools will evolve: 
information gathering and processing agents.
This paper describes {\sc wave} 
(Web Analysis and Visualization Environment), 
a 3D interface for World-Wide Web
information visualization and browsing.
It uses the mathematical theory of concept analysis 
to conceptually cluster objects, 
and 
to create a three-dimensional layout of information nodes.
So-called ``conceptual scales'' for attributes,
such as location, title, keywords, topic, size, or modification time,
provide a formal mechanism 
that automatically classifies and categorizes documents,
creating a conceptual information space. 
A visualization shell serves as 
an ergonomically sound user interface
for exploring this information space.
\end{abstract}


\section{Introduction}
\label{INTRODUCTION}

The {\bf World-Wide Web} has gained its amazing popularity 
through the availability of ``point and shoot'' browsing tools like {\bf Mosaic}.
They provide access to information sources all over the globe 
via a simple, easy to use graphical interface.
The information space that is available to Web users is enormous,
and now that editing tools offer word processor like ease-of-use,
its growth rate will certainly accelerate.
Thus,
the user will literally drown in an ocean of information 
--- the experience of getting ``lost in hyperspace'' 
is probably already familiar to most Web citizens.


In order to provide Web users with a facility for
resource discovery, so called {\it meta index\/}
servers have been set up which maintain lists of
references to other servers and resources. There
is however the problem that these reference become
stale whenever documents move or are deleted.
Thus, some server maintainers have automated this
process by retrieving and parsing documents in
regular intervals. A popular example is the
searchable CUI W3 catalog\footnote{
  \protect\verb~<http://cui\_www.unige.ch/w3catalog>~
} run by the University of Geneva.
Another approach is
Martijn Koster's {\sc aliweb} \cite{Kos94}, which creates
an archie-like indexing facility: the {\sc aliweb}
server regularly retrieves index files from other
servers and combines them into a searchable database.
An interesting implementation of client based searches
is the {\it fish search\/} \cite{BrPo94} extension to Mosaic which
was developed at the Eindhoven University of Technology
in the Netherlands: It enhances the Mosaic Web browser
with a functionality for robotic searches on a remote
archive of limited depth 
(section~\ref{GATHER} will explain the term
robotic search).
Use of a local cache avoids multiple
accesses to the same document.

These tools allow for topic searches based on keywords,
and return a list of resource references which meet
a given condition. However, since the result of
a keyword search is often too large to be handled
conveniently, there is a need for a more refined
method of categorizing and managing information.
By employing autonomous agents that retrieve and
automatically analyze information from the Web,
a sophisticated system for information retrieval
and management can be created.

The process of automated document analysis can
be divided into three phases:
\begin{itemize}
\itemsep 0ex plus 0ex minus 0ex 
\parsep 0ex plus 0ex minus 0ex 
\item acquisition of raw data
\item automatic analysis and classification
\item visualization and interactive browsing
\end{itemize}

The following sections will discuss in detail how
these phases are implemented in {\sc wave}.


\section{Information gathering}
\label{GATHER}

An information analyzer needs to gather raw data from
remote information repositories.
The straightforward approach
is to use an automatic spider program to retrieve documents from
a remote host. A Web Spider program (also referred to as a
Web Wanderer or Web Robot) is a routine that recursively
retrieves documents.
The term ``web wanderer'' is a bit misleading: these
programs do not cross machine boundaries, but are
executed on the local machine. In order to search a 
World Wide Web archive, they have to transfer a whole 
document tree over the network.

Of course, these programs are very bandwidth intensive and
have quite an impact on the servers processing resources.
The bandwidth requirements can be reduced by having a robot
program access a local Proxy cache \cite{LuAl94} instead of directly
connecting to a remote site. Especially when re-analyzing
a remote document repository, use of cache mechanisms
significantly reduces the required bandwidth by limiting
document transfers to those files which actually have
changed in the meantime.

Another possibility is the employment of automatically
created indices. Document indexing tools like
{\sc waisindex} or
ICE allow for making keyword based topic searches, but
can also be used to provide a well defined interface for
operations like returning a list of document URLs in the 
archive together with attributes like size, date of last 
modification etc.

And there is a third mechanism possible which seems
a bit exotic at the first glance: software agents
capable of crossing the border between machines and
executing their search task on a the remote target.
Such an agent could
truly be called a ``web wanderer'' --- since the search
is being executed on the target machine, only the
results have to be transferred back, thus greatly
reducing the required bandwidth. Such agents can be
implemented as scripts, which are being interpreted
in a save environment by a ``script engine'' located on
the remote host. 

The next section will show how 
the raw data about Web documents,
which has been gathered by devices such as described here,
is preprocessed in order to generate conceptual scales to be used
for the classification of Web documents.


\section{Interpretation and Classification}

Ideally, 
an information processing agent should be able to actually read 
and to a certain degree ``understand'' a document.
Although recent developments by artificial intelligence researchers
in the area of machine translation are promising,
current technology is based upon 
the use of heuristic methods for the classification of documents.

In {\sc wave}
we combine concepts, techniques, and processes
from both
traditional Library Science (cataloging and classification)
\cite{Wy80}
and
the relatively new discipline of Concept Analysis
\cite{GaWi89}.
Collections of Web documents should be arranged according to some system,
and such an arrangement is referred to as a classification.
The purpose of classification is 
to make each document readily available to the user.
The goal of {\sc wave}
is the arrangement of Web documents into conceptual classes
which exhibit all of the resemblances and differences
essential to their full comprehension by the user.

\subsection{Conceptual Classes}

A {\em conceptual class\/} consists of any group of entities or objects
exhibiting one or more common characteristics, traits or attributes.
A characteristic is a conceptualized attribute
by which classes may be 
identified and separated into a conceptual hierarchy, and
further subdivided (specialized) by the facets of
topic, form, location, chronology, etc.
The ``has'' relationship between objects and attributes
is represented as a binary relation called a formal context.
A {\em formal context\/} is a triple $\triple{G}{M}{I}$
consisting of two sets $G$ and $M$
and a binary incidence relation $I \subseteq \product{G}{M}$ between $G$ and $M$.
Intuitively, 
the elements of $G$ are thought of as entities or objects,
the elements of $M$ are thought of as properties, characteristics or attributes
that the objects might have,
and $g{I}m$ asserts that ``object $g$ has attribute $m$.''
In many contexts appropriate for Web documents,
the objects are documents
and the attributes are any interesting properties of those documents.

The definition of a conceptual class must involve:
the common attributes,
which are encoded in the superordinate 
(next higher and more general class),
and 
the distinguishing attributes,
which differentiate the defined concept from the superordinate.
Conceptual classes are logically characterized by their extension and intension.
\begin{itemize}
	\item The {\em extension\/} of a class is 
		the aggregate of entities or objects
		which it includes or denotes.
	\item The {\em intension\/} of a class is 
		the sum of its unique characteristics, traits or attributes,
		which, taken together,
		imply the concept signified by the conceptual class.
\end{itemize}
The intent should contain precisely those attributes 
shared by all objects in the extent,
and vice-versa,
the extent should contain precisely those objects 
sharing all attributes in the intent.
Clearly the terms ``extension'' and ``intension'' are reciprocally dependent.
They complement each other by 
reciprocally deliminating concepts and explicating definitions.
A conceptual class will consist of such an extent/intent pair.

The process of subordination of conceptual classes and collocation of objects
exhibits a natural order,
proceeding top-down
from the more general classes with larger extension and smaller intension
to the more specialized classes with smaller extension and larger intension.
This order is called generalization-specialization.
One class is more specialized (and less general) than another class,
when its intent contains the other's intent,
or equivalently,
when the opposite ordering on extents occurs.
Conceptual classes with this generalization-specialization ordering
form a class hierarchy for the formal context.
Knowledge is here represented as the hierarchical structure 
known as a complete lattice,
and called the {\em concept lattice\/} of the formal context.

The join of a collection of conceptual classes
represents the common attributes or shared characteristics of the classes.
The bottom of the conceptual hierarchy (the empty join)
represents the most specific class 
whose intent consists of all attributes
and whose extent is often empty.
The meet of a collection of conceptual classes
represents the conjunction of all the attributes of the classes.
The top of the conceptual hierarchy (the empty meet)
represents the universal class whose extent consists of all objects.
The entire conceptual class hierarchy is implicitly specified
by the ``has'' relationship of the formal context.
However,
part of the hierarchy of conceptual classes 
could also be explicitly specified 
via the following top-down process \cite{Wy80}.
\begin{itemize}
	\item {\bf Initialization:} 
		The main top-level attribute classes are specified.
		These are meet-irreducible classes,
		meaning that they cannot be expressed
		as the meet of other more general classes.
	\item {\bf Iteration:} 
		Any collection of (super)classes can be specialized
		by the specification of differentiating attributes,
		thus producing subclasses.
		Each such differentiated (sub)class 
		is subordinate to 
		every (super)class in the collection.
	\item {\bf Termination:}
		Continue until further specialization and differentiation
		is either impossible or impractical.
\end{itemize}

\subsection{Conceptual Scaling}

The general perspective of Concept Analysis fits closely with the 
``information workspace'' paradigm, as described in \cite{RCM93}.
The interpretive act (called conceptual scaling in Concept Analysis) is 
all-important, since this is the way the user makes sense of his world of data.
Interpretation can automatically and implicitly define 
the classification and categorization of data objects, 
and hence is more fundamental than classification.
Interpretation is effected by conceptual scales.
A natural approach 
toward the enrichment of interpretation and classification
uses ideas from fuzzy sets and rough sets \cite{Ke94}.

\begin{figure}[htb]
\begin{center}
\setlength{\unitlength}{0.8pt}
\newcommand{\puttext}[3]{\put(#1,#2){{\mbox{\tiny$#3$\normalsize}}}}
\newcommand{\putdisk}[3]{\put(#1,#2){\circle*{#3}}}
\begin{tabular}[t]{|c|}\hline
	{\bf naming scheme scale} \\ \hline\hline
\begin{picture}(200,200)
\put(20,25){\begin{picture}(150,150)
	\puttext{0}{0}{{\bf }}
	\putdisk{75}{150}{7}			
	\putdisk{0}{75}{7}			
	\puttext{5}{80}{\mbox{\bf file}}	
	\puttext{5}{70}{\mbox{scheme}}		
	\puttext{10}{63}{=\mbox{file}}		
	\putdisk{50}{75}{7}			
	\puttext{55}{80}{\mbox{\bf gopher}}	
	\puttext{55}{70}{\mbox{scheme}}		
	\puttext{60}{63}{=\mbox{gopher}}	
	\putdisk{100}{75}{7}			
	\puttext{105}{80}{\mbox{\bf news}}	
	\puttext{105}{70}{\mbox{scheme}}	
	\puttext{110}{63}{=\mbox{news}}		
	\putdisk{150}{75}{7}			
	\puttext{155}{80}{\mbox{\bf http}}	
	\puttext{155}{70}{\mbox{scheme}}	
	\puttext{160}{63}{=\mbox{http}}		
	\putdisk{75}{0}{7}			
	\put(0,75){\line(1,1){75}}		
	\put(50,75){\line(1,3){25}}		
	\put(100,75){\line(-1,3){25}}		
	\put(150,75){\line(-1,1){75}}		
	\put(75,0){\line(-1,1){75}}		
	\put(75,0){\line(-1,3){25}}		
	\put(75,0){\line(1,3){25}}		
	\put(75,0){\line(1,1){75}}		
\end{picture}}
\end{picture}
\\ \hline
\end{tabular}
\end{center}
\caption{{\bf Naming Scheme scale}}
\label{scheme:scale}
\end{figure}
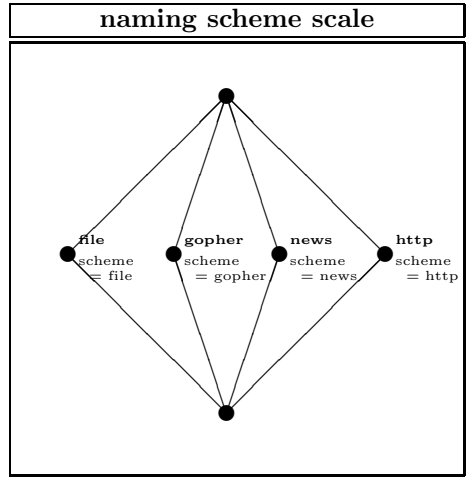

A conceptual scale is a single isolated trait, property, or use,
which is distinct from other characteristics.
It is a kind of filter for a single conceptual dimension of data.
We identify the notion of conceptual scale from Concept Analysis
with the notion of facet from Library Science.
The collection of conceptual scales in use (often chosen by the user)
defines a certain view of the universe of Web documents.
Here we list some examples of Web-related conceptual scales.
\begin{enumerate}
\item The {\it location} of {\sc html} documents could be scaled 
	either nationally, geographically, net-wise, or other.
	This example discusses the standard hierarchical approach to
	geographical facets or scales on the Web.
	This hierarchy,
	as many other hierarchical scales (without instances),
	has a common set of attributes and objects,
	resulting in a square incidence matrix in its formal context.
	When instances are included,
	the context will no longer be square.
	The multi-valued attributes 
	for the geographical scale in Table~\ref{geog:attributes}
	have the functional dependencies
	\[ 
	  \mbox{city} \Rightarrow \mbox{state},\;\;
	  \mbox{state} \Rightarrow \mbox{country},\;\;
	  \mbox{country} \Rightarrow \mbox{continent},\;\;
	  \mbox{continent} \Rightarrow \mbox{hemisphere} 
	\]
	These multi-valued attributes will each individually be nominally scaled,
	providing the notion of levels in the overall hierarchical scale.
	This hierarchical scale is assembled by instantiating the functional dependencies.
	The pairings of inclusion,
	which are placed in an incidence matrix file (\verb|*.tbl|),
	are listed in Table~\ref{geog:scale}.
	If these attributes (and objects) are sorted by level,
	from larger to smaller areas,
	the resulting incidence matrix will be block lower triangular.
	If we partition the attributes by level,
	resulting in a collection of conceptual scales,
	then by using nested line diagrams
	we can abstractly visualize the hierarchy.
	By dropping attributes right-to-left,
	we can effect a kind of abstraction-by-restriction.
\item The {\sc url}s of Web documents can be scaled.
	The information contained in {\sc url}s,
	and more completely in {\sc ur*}s
	({\sc uri}s, {\sc urn}s, {\sc url}s, and especially {\sc urc}s),
	corresponds to the bibliographic records in library catalogs of Library Science.
	\begin{enumerate}
	\item The {\sc url} naming scheme component
		is scaled in Figure~\ref{scheme:scale}.
	\item Instantiation of the geographical scale with respect to 
		the hostname component of {\sc url}s
		is a kind of morphism of contexts
		${\rm location} \colon {\rm url} \longrightarrow {\rm city}$.
		For example,
		an instance of this is
		{\tt http://www.cern.ch/} $\longmapsto$ {\tt www.cern.ch} 
                                          $\longmapsto$ {\tt geneva}.
		When this context morphism 
		is combined with the geographical scale mentioned above,
		the {\sc url} hostname component is scaled hierarchically by level.
	\item The {\sc url} path component can be conceptually scaled.
		Although it does not (yet) contain explicit semantics,
		and may not even map to a physical directory structure,
		it usually still reflects a hierarchy of documents.
		It is safe to assume that documents residing
		in the same {\sc url} ``directory'' are related in some way,
		and that the  hierarchy in the {\sc url}s reflects a
		logical, hierarchical clustering of documents.
	\end{enumerate}
\item In addition to {\sc url} scales for documents, 
	a user may also be interested in 
	{\it size\/} scales, {\it form\/} scales, {\it time\/} scales, etc.
	But most important
	from the standpoint of semantics
	will be the various {\it subject\/}, {\it content\/}, or {\it topic} scales 
	(compare the subject headings for a classification system 
	such as Dewey Decimal, Library of Congress, Bliss, Colon, etc.).
\end{enumerate}

\subsection{Apposition of Conceptual Scales}

Conceptual scales can be combined in various ways.
However,
the most useful way for the classification of Web documents
is by {\em apposition\/} of conceptual scales \cite{GaWi89}.
Apposition is a kind of product or conjunction of conceptual scales.
It combines 
the various relevant and purposeful conceptual data dimensions
into a clear, unambiguous aggregate
which allows for several visual abstractions called nested line diagrams.

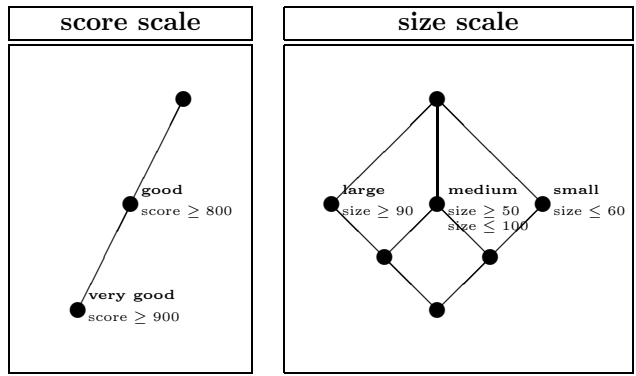
\begin{figure}[htb]
\begin{center}
\setlength{\unitlength}{0.8pt}
\newcommand{\puttext}[3]{\put(#1,#2){{\mbox{\tiny$#3$\normalsize}}}}
\newcommand{\putdisk}[3]{\put(#1,#2){\circle*{#3}}}
\begin{tabular}{cc}
\begin{tabular}[t]{|c|}\hline
	{\bf score scale} \\ \hline\hline
\begin{picture}(100,150)
\put(25,25){\begin{picture}(50,100)
	\puttext{0}{0}{{\bf }}
	\putdisk{50}{100}{7}			
	\putdisk{25}{50}{7}			
	\puttext{30}{55}{\mbox{\bf good}}	
	\puttext{30}{45}{\mbox{score} \geq 800}	
	\putdisk{0}{0}{7}			
	\puttext{5}{5}{\mbox{\bf very good}}	
	\puttext{5}{-5}{\mbox{score} \geq 900}	
	\put(0,0){\line(1,2){25}}		
	\put(25,50){\line(1,2){25}}		
\end{picture}}
\end{picture}
\\ \hline
\end{tabular}
&
\begin{tabular}[t]{|c|}\hline
	{\bf size scale} \\ \hline\hline
\begin{picture}(150,150)
\put(15,25){\begin{picture}(100,100)
	\putdisk{50}{100}{7}			
	\putdisk{0}{50}{7}			
	\puttext{5}{55}{\mbox{\bf large}}	
	\puttext{5}{45}{\mbox{size} \geq 90}	
	\putdisk{50}{50}{7}			
	\puttext{55}{55}{\mbox{\bf medium}}	
	\puttext{55}{45}{\mbox{size} \geq 50}	
	\puttext{55}{38}{\mbox{size} \leq 100}	
	\putdisk{25}{25}{7}			
	\putdisk{100}{50}{7}			
	\puttext{105}{55}{\mbox{\bf small}}	
	\puttext{105}{45}{\mbox{size} \leq 60}	
	\putdisk{75}{25}{7}			
	\putdisk{50}{0}{7}			
	\put(0,50){\line(1,1){50}}		
	\put(50,50){\line(0,1){50}}		
	\put(100,50){\line(-1,1){50}}		
	\put(25,25){\line(-1,1){25}}		
	\put(25,25){\line(1,1){25}}		
	\put(75,25){\line(-1,1){25}}		
	\put(75,25){\line(1,1){25}}		
	\put(50,0){\line(-1,1){25}}		
	\put(50,0){\line(1,1){25}}		
\end{picture}}
\end{picture}
\\ \hline
\end{tabular}
\end{tabular}
\end{center}
\caption{{\bf score \& size scales: Ethnic Cooking}}
\label{scales}
\end{figure}

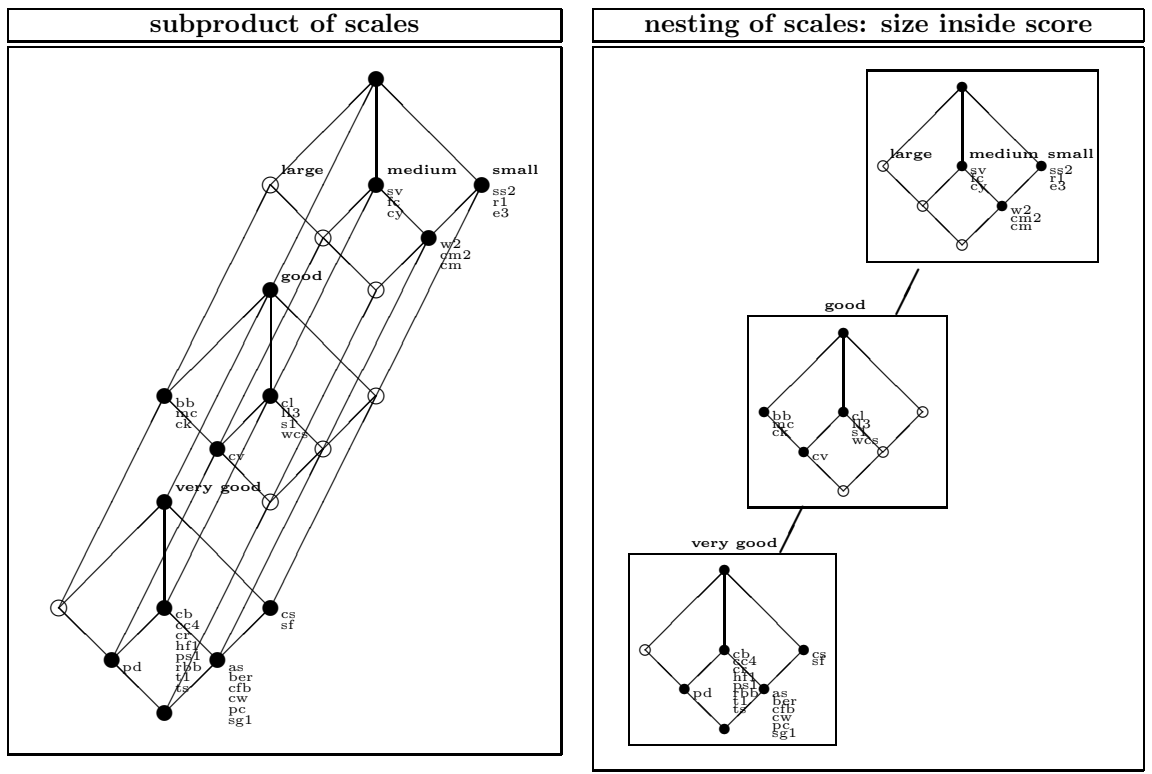
\begin{figure}[thb]
\begin{center}
\begin{tabular}{cc}

\begin{tabular}[t]{|c|}\hline
{\bf subproduct of scales}
\\ \hline\hline
\setlength{\unitlength}{0.8pt}
\newcommand{\puttext}[3]{\put(#1,#2){{\mbox{\tiny$#3$\normalsize}}}}
\newcommand{\putcirc}[3]{\thinlines\put(#1,#2){\circle{#3}}}
\newcommand{\putdisk}[3]{\put(#1,#2){\circle*{#3}}}
\begin{picture}(230,330)(0,-15)
	\puttext{0}{0}{{\bf }}
\put(100,200){\begin{picture}(100,100)
	\putdisk{50}{100}{7}		
	\putcirc{0}{50}{7}			
	\puttext{5}{55}{{\bf large}}	
	\putdisk{50}{50}{7}			
	\puttext{55}{55}{{\bf medium}}
	\puttext{55}{45}{{\rm sv}}	
	\puttext{55}{40}{{\rm fc}}	
	\puttext{55}{35}{{\rm cy}}	
	\putcirc{25}{25}{7}			
	\putdisk{100}{50}{7}		
	\puttext{105}{55}{{\bf small}}
	\puttext{105}{45}{{\rm ss2}}	
	\puttext{105}{40}{{\rm r1}}	
	\puttext{105}{35}{{\rm e3}}	
	\putdisk{75}{25}{7}			
	\puttext{80}{20}{{\rm w2}}	
	\puttext{80}{15}{{\rm cm2}}	
	\puttext{80}{10}{{\rm cm}}	
	\putcirc{50}{0}{7}			
	\put(0,50){\line(1,1){50}}	
	\put(50,50){\line(0,1){50}}	
	\put(100,50){\line(-1,1){50}}	
	\put(25,25){\line(-1,1){25}}	
	\put(25,25){\line(1,1){25}}	
	\put(75,25){\line(-1,1){25}}	
	\put(75,25){\line(1,1){25}}	
	\put(50,0){\line(-1,1){25}}	
	\put(50,0){\line(1,1){25}}	
\end{picture}}
\put(50,100){\begin{picture}(100,100)
	\putdisk{50}{100}{7}		
	\puttext{55}{105}{\mbox{\bf good}}	
	\putdisk{0}{50}{7}			
	\puttext{5}{45}{{\rm bb}}	
	\puttext{5}{40}{{\rm mc}}	
	\puttext{5}{35}{{\rm ck}}	
	\putdisk{50}{50}{7}			
	\puttext{55}{45}{{\rm cl}}	
	\puttext{55}{40}{{\rm ll3}}	
	\puttext{55}{35}{{\rm s1}}	
	\puttext{55}{30}{{\rm wcs}}	
	\putdisk{25}{25}{7}			
	\puttext{30}{20}{{\rm cv}}	
	\putcirc{100}{50}{7}		
	\putcirc{75}{25}{7}			
	\putcirc{50}{0}{7}			
	\put(0,50){\line(1,1){50}}	
	\put(50,50){\line(0,1){50}}	
	\put(100,50){\line(-1,1){50}}	
	\put(25,25){\line(-1,1){25}}	
	\put(25,25){\line(1,1){25}}	
	\put(75,25){\line(-1,1){25}}	
	\put(75,25){\line(1,1){25}}	
	\put(50,0){\line(-1,1){25}}	
	\put(50,0){\line(1,1){25}}	
\end{picture}}
\put(0,0){\begin{picture}(100,100)
	\putdisk{50}{100}{7}		
	\puttext{55}{105}{\mbox{\bf very good}}	
	\putcirc{0}{50}{7}			
	\putdisk{50}{50}{7}			
	\puttext{55}{45}{{\rm cb}}	
	\puttext{55}{40}{{\rm cc4}}	
	\puttext{55}{35}{{\rm cr}}	
	\puttext{55}{30}{{\rm hf1}}	
	\puttext{55}{25}{{\rm ps1}}	
	\puttext{55}{20}{{\rm rbb}}	
	\puttext{55}{15}{{\rm t1}}	
	\puttext{55}{10}{{\rm ts}}	
	\putdisk{25}{25}{7}			
	\puttext{30}{20}{{\rm pd}}	
	\putdisk{100}{50}{7}		
	\puttext{105}{45}{{\rm cs}}	
	\puttext{105}{40}{{\rm sf}}	
	\putdisk{75}{25}{7}			
	\puttext{80}{20}{{\rm as}}	
	\puttext{80}{15}{{\rm ber}}	
	\puttext{80}{10}{{\rm cfb}}	
	\puttext{80}{5}{{\rm cw}}	
	\puttext{80}{0}{{\rm pc}}	
	\puttext{80}{-5}{{\rm sg1}}	
	\putdisk{50}{0}{7}			
	\put(0,50){\line(1,1){50}}	
	\put(50,50){\line(0,1){50}}	
	\put(100,50){\line(-1,1){50}}	
	\put(25,25){\line(-1,1){25}}	
	\put(25,25){\line(1,1){25}}	
	\put(75,25){\line(-1,1){25}}	
	\put(75,25){\line(1,1){25}}	
	\put(50,0){\line(-1,1){25}}	
	\put(50,0){\line(1,1){25}}	
\end{picture}}
	\put(100,200){\line(1,2){50}}	
	\put(50,150){\line(1,2){50}}	
	\put(100,150){\line(1,2){50}}	
	\put(150,150){\line(1,2){50}}	
	\put(125,125){\line(1,2){50}}	
	\put(75,125){\line(1,2){50}}	
	\put(100,100){\line(1,2){50}}	
	\put(50,100){\line(1,2){50}}	
	\put(0,50){\line(1,2){50}}	
	\put(50,50){\line(1,2){50}}	
	\put(100,50){\line(1,2){50}}	
	\put(75,25){\line(1,2){50}}	
	\put(25,25){\line(1,2){50}}	
	\put(50,0){\line(1,2){50}}	
\end{picture}
\\ \hline
\end{tabular}
&	
\begin{tabular}[t]{|c|}\hline
{\bf nesting of scales: size inside score}
\\ \hline\hline
\setlength{\unitlength}{0.6pt}
\newcommand{\puttext}[3]{\put(#1,#2){{\mbox{\tiny$#3$\normalsize}}}}
\newcommand{\putcirc}[3]{\thinlines\put(#1,#2){\circle{#3}}}
\newcommand{\putdisk}[3]{\put(#1,#2){\circle*{#3}}}
\begin{picture}(305,450)(5,-20)
	\puttext{0}{0}{{\bf }}
\put(145,300){\fbox{\begin{picture}(135,110)(-5,-5)
	\putdisk{50}{100}{7}		
	\putcirc{0}{50}{7}			
	\puttext{4}{55}{{\bf large}}	
	\putdisk{50}{50}{7}			
	\puttext{54}{55}{{\bf medium}}
	\puttext{55}{45}{{\rm sv}}	
	\puttext{55}{40}{{\rm fc}}	
	\puttext{55}{35}{{\rm cy}}	
	\putcirc{25}{25}{7}			
	\putdisk{100}{50}{7}		
	\puttext{104}{55}{{\bf small}}
	\puttext{105}{45}{{\rm ss2}}	
	\puttext{105}{40}{{\rm r1}}	
	\puttext{105}{35}{{\rm e3}}	
	\putdisk{75}{25}{7}			
	\puttext{80}{20}{{\rm w2}}	
	\puttext{80}{15}{{\rm cm2}}	
	\puttext{80}{10}{{\rm cm}}	
	\putcirc{50}{0}{7}			
	\put(0,50){\line(1,1){50}}	
	\put(50,50){\line(0,1){50}}	
	\put(100,50){\line(-1,1){50}}	
	\put(25,25){\line(-1,1){25}}	
	\put(25,25){\line(1,1){25}}	
	\put(75,25){\line(-1,1){25}}	
	\put(75,25){\line(1,1){25}}	
	\put(50,0){\line(-1,1){25}}	
	\put(50,0){\line(1,1){25}}	
\end{picture}}}
\put(70,145){\fbox{\begin{picture}(115,110)(-5,-5)
	\puttext{38}{115}{\mbox{\bf good}}	
	\putdisk{50}{100}{7}		
	\putdisk{0}{50}{7}			
	\puttext{5}{45}{{\rm bb}}	
	\puttext{5}{40}{{\rm mc}}	
	\puttext{5}{35}{{\rm ck}}	
	\putdisk{50}{50}{7}			
	\puttext{55}{45}{{\rm cl}}	
	\puttext{55}{40}{{\rm ll3}}	
	\puttext{55}{35}{{\rm s1}}	
	\puttext{55}{30}{{\rm wcs}}	
	\putdisk{25}{25}{7}			
	\puttext{30}{20}{{\rm cv}}	
	\putcirc{100}{50}{7}		
	\putcirc{75}{25}{7}			
	\putcirc{50}{0}{7}			
	\put(0,50){\line(1,1){50}}	
	\put(50,50){\line(0,1){50}}	
	\put(100,50){\line(-1,1){50}}	
	\put(25,25){\line(-1,1){25}}	
	\put(25,25){\line(1,1){25}}	
	\put(75,25){\line(-1,1){25}}	
	\put(75,25){\line(1,1){25}}	
	\put(50,0){\line(-1,1){25}}	
	\put(50,0){\line(1,1){25}}	
\end{picture}}}
\put(-5,-5){\fbox{\begin{picture}(120,110)(-5,-5)
	\puttext{29}{115}{\mbox{\bf very good}}	
	\putdisk{50}{100}{7}		
	\putcirc{0}{50}{7}			
	\putdisk{50}{50}{7}			
	\puttext{55}{45}{{\rm cb}}	
	\puttext{55}{40}{{\rm cc4}}	
	\puttext{55}{35}{{\rm cr}}	
	\puttext{55}{30}{{\rm hf1}}	
	\puttext{55}{25}{{\rm ps1}}	
	\puttext{55}{20}{{\rm rbb}}	
	\puttext{55}{15}{{\rm t1}}	
	\puttext{55}{10}{{\rm ts}}	
	\putdisk{25}{25}{7}			
	\puttext{30}{20}{{\rm pd}}	
	\putdisk{100}{50}{7}		
	\puttext{105}{45}{{\rm cs}}	
	\puttext{105}{40}{{\rm sf}}	
	\putdisk{75}{25}{7}			
	\puttext{80}{20}{{\rm as}}	
	\puttext{80}{15}{{\rm ber}}	
	\puttext{80}{10}{{\rm cfb}}	
	\puttext{80}{5}{{\rm cw}}	
	\puttext{80}{0}{{\rm pc}}	
	\puttext{80}{-5}{{\rm sg1}}	
	\putdisk{50}{0}{7}			
	\put(0,50){\line(1,1){50}}	
	\put(50,50){\line(0,1){50}}	
	\put(100,50){\line(-1,1){50}}	
	\put(25,25){\line(-1,1){25}}	
	\put(25,25){\line(1,1){25}}	
	\put(75,25){\line(-1,1){25}}	
	\put(75,25){\line(1,1){25}}	
	\put(50,0){\line(-1,1){25}}	
	\put(50,0){\line(1,1){25}}	
\end{picture}}}
\thicklines
	\put(91,112){\line(1,2){14}}	
	\put(164,262){\line(1,2){14}}	
\thinlines
\end{picture}
\\ \hline
\end{tabular}

\end{tabular}
\end{center}
\caption{{\bf Conceptual Scaling of a search result: Ethnic Cooking}}
\label{wais}
\end{figure}

For an example of apposition of conceptual scales,
consider Table~\ref{gfro}.
Table~\ref{gfro} shows the results of a {\sc waisindex} search
in the library \verb|usenet-cookbook.src|
using the set of keywords ``garlic'', ``fish'', 
``rice'', and ``onion''.
The documents are recipes,
the score is the {\sc wais} relative score,
and
the size is the number of lines in the recipe.
This example was discussed by 
Ed Krol \cite{Kr94}
as an example of {\sc wais} search.
In Figure~\ref{scales}
the scores are scaled in an ordinal scale
and
and the document size is scaled with an interordinal scale.
Figure~\ref{wais} uses 
apposition of the score and size scales in Figure~\ref{scales}
with the idea of nesting
to visualize in a line diagram (concept lattice)
the conceptual scaling of the results of a {\sc waisindex}
search in Table~\ref{gfro} using the scales in Figure~\ref{scales}.

On the left side of Figure~\ref{wais} is displayed 
the subproduct of the score and size scale:
the collection of all 21 nodes (filled or unfilled)
represents the product of the score scale and the size scale in Figure~\ref{scales};
the collection of 14 filled nodes represents the sublattice
which results from instantiation of the product with respect to
the raw data in Table~\ref{gfro}
(filter the document information in Table~\ref{gfro}
from the library \verb|usenet-cookbook.src|
through the conceptual scales).

On the right side in Figure~\ref{wais} is displayed 
the nesting of the size scale inside of the score scale.
This is a kind of visual abstraction.
In general,
the inner instatiated scale is substituted into the outer instantiated scale
by intersecting object sets.
The outer scale is viewed as a rough approximation to the nested scales.
Nesting can occur to any level,
with the number of levels corresponding to 
the number of component scales in apposition.
The order of nesting could correspond to the rank order of attributes.
In {\sc wave} a 3D version of nested line diagrams will be developed.

\subsection{Kinds of Conceptual Scales}

Based upon their structural properties,
conceptual scales can be classified into several kinds.
Conceptual scales (the mechanism for interpretation) can be implemented
as autonomous interpretative agents.
Conceptual scales can themselves be scaled by kind
--- interpretative meta-agents can control interpretative agents.
From a utilitarian standpoint
some of the more important kinds of scales are the following.
\begin{itemize}
	\item {\bf Nominal scales} represent partition and independence.
		The {\sc url} naming scheme scale
		in Figure~\ref{scheme:scale} is nominally scaled.
	\item {\bf Ordinal scales} (one-dimensional) represent ranking.
		The score scale for the Ethnic Cooking search example 
		in Figure~\ref{scales} is ordinally scaled.
		Figure~\ref{wais} shows at a glance that
		``CUBAN-BEANS'' is ranked higher than ``CAJUN-LAMB''
		with respect to score.
	\item {\bf Interordinal scales} represent betweenness.
		The size scale for the Ethnic Cooking search example 
		in Figure~\ref{scales} is interordinally scaled.
		Again,
		Figure~\ref{wais} shows at a glance that
		the ``CHICKEN-VINDAL'' recipe is between
		the ``BOUILLABAISSE'' recipe and the ``CAJUN-LAMB'' recipe
		in size.
	\item {\bf Hierarchical scales} represent, of course, hierarchical
		information;
		either in the single-inheritance case of a tree hierarchy,
		or in the multi-inheritance case.
		The geographical scale of server site addresses
		in Table~\ref{geog:attributes} and Table~\ref{geog:scale}
		is hierarchically scaled by level.
\end{itemize}

\subsection{Types of Classification}

Referential classification \cite{Wy80} is a pragmatic and empirical system
in which objects are related
with reference to
a chosen collection of conceptual scales.
In referential classification,
various external relations, user preferences, and the environment,
are all important to the act of interpretation and classification.
Any Web document may be meaningful in any number of different relationships,
depending upon the immediate purpose of the user.
To support this flexible interpretation,
classification in {\sc wave}
allows the user to define a new view 
which is based upon a different collection of conceptual scales.
So {\sc wave} is a referential classification system.

A faceted classication scheme
tends to restrict explicit designations to single, unsubdivided classes.
These are called meet irreducible conceptual classes in the class hierarchy.
This list of designations can be identified with attribute names,
since all meet irreducible classes are labeled by such names.
Faceted classification schemes rely upon synthesis
--- the combining of various facets (conceptual scales) 
for the specification and construction of conceptual classes.
The Colon classification scheme of Ranganathan is one example 
of faceted classification from Library Science.
The {\sc wave} system is another example
of faceted classification.

\subsection{The Process of Interpretation and Classification}

The {\sc wave} process of classifying a Web document
with conceptual scales is an act of interpretation.
This process,
which constructs classes and embeds them in conceptual class hierarchies,
involves the synthesis of conceptual classes using conceptual scales.
It consists of the following steps.
\renewcommand{\theenumi}{\Roman{enumi}}
\renewcommand{\theenumii}{\roman{enumii}}
\begin{center}
\begin{enumerate}
	\item {\bf Analysis:} 
	\begin{enumerate}
		\item Gather into a cache all relevant information about a document:
			physical information, content, form, etc.
		\item Break down the document information into its component parts.
			Resolve the information into ``atomic units'';
			usually nouns and descriptive adjectives, or
			linguistic variables and linguistic values.
	\end{enumerate}
	\item {\bf Synthesis}
	\begin{enumerate}
		\item Identify the appropriate conceptual scale or facet
			for each atomic unit of information.
		\item Filter the document information through the conceptual scales.
		\item Construct the conceptual class of the document
			by forming the lattice meet in the class hierarchy
			of the collection of attribute conceptual classes
			of the faceted or scaled document information.
	\end{enumerate}
\end{enumerate}
\end{center}
\renewcommand{\theenumi}{\arabic{enumi}}
\renewcommand{\theenumii}{\alph{enumii}}
Through the use of conceptual scaling
the {\sc wave} system builds up and synthesizes
the features of Web documents
---
such as purpose, form, location, estimated access time, size, time of last revision, etc.
---
into a conceptual structure,
a systematized and orderly whole.


\scriptsize
\normalsize


\scriptsize
\begin{table}[htb]
\begin{center}
\begin{tabular}{|l|l|}\hline
	\multicolumn{1}{|c|}{{\bf attribute}}
		& \multicolumn{1}{c|}{{\bf domain}} \\ \hline\hline
	hemisphere
		& $\{ \mbox{eastern}, \mbox{western} \}$ \\ \hline
	continent
		& $\{ \cdots , \mbox{europe}, 
		      \cdots , \mbox{america:north}, \cdots \}$ \\ \hline
	country
		& $\{ \cdots , \mbox{britain}, \cdots , \mbox{germany}, 
			 \cdots , \mbox{united\_states}, \cdots \}$ \\ \hline
	state, province, land
		& $\{ \cdots , \mbox{scotland}, \cdots , \mbox{hesse}, 
		      \cdots , \mbox{alaska}, \cdots \}$ \\ \hline
	city
		& $\{ \cdots , \mbox{edinburgh}, \cdots , \mbox{frankfurt}, 
			 \cdots , \mbox{anchorage}, \cdots \}$ \\ \hline
\end{tabular}
\end{center}
\caption{{\bf The multi-valued attributes for the Geographical scale}}
\label{geog:attributes}
\end{table}
\normalsize

\scriptsize
\begin{table}[htb]
\begin{center}
\fbox{\begin{tabular}{ll}
  \multicolumn{2}{c}{\ldots} \\
  continent:asia    & hemisphere:eastern \\
  \multicolumn{2}{c}{\ldots} \\
  country:britain   & continent:europe \\
  \multicolumn{2}{c}{\ldots} \\
  state:alaska      & country:united\_states \\
  \multicolumn{2}{c}{\ldots} \\
  city:frankfurt    & land:hessen \\
  \multicolumn{2}{c}{\ldots}
\end{tabular}}
\end{center}
\caption{{\bf The File} 
         {\tt geographical.scale.tbl} 
         {\bf for the Geographical Scale}}
\label{geog:scale}
\end{table}
\normalsize


\scriptsize
\begin{table}[htb]
\begin{center}
\begin{tabular}{cc}
\begin{tabular}[t]{|rl||r|r|}\hline
	\multicolumn{2}{|c||}{\bf document}	
		& \multicolumn{1}{|c}{\bf score}	
			& \multicolumn{1}{|c|}{\bf size} \\ \hline\hline
	SARDINE-FRY	&(sf)	& 1000	& 47 \\ \hline
	CURRIED-RICE	&(cr)	& 993	& 67 \\ \hline
	RICE-BEAN-BAKE	&(rbb)	& 986	& 61 \\ \hline
	HOT-FANNY-1	&(hf)	& 972	& 64 \\ \hline
	STROGANOFF-1	&(sg1)	& 965	& 56 \\ \hline
	BLACK-EYE-RICE	&(ber)	& 958	& 57 \\ \hline
	TORTILLA-SOUP	&(ts)	& 943	& 74 \\ \hline
	CABBAGE-SALAD	&(cs)	& 943	& 45 \\ \hline
	PONCIT		&(pc)	& 936	& 54 \\ \hline
	PEANUT-SAUCE-1	&(ps1)	& 936	& 62 \\ \hline
	CUBAN-BEANS	&(cb)	& 936	& 67 \\ \hline
	AFRICAN-STEW	&(as)	& 936	& 56 \\ \hline
	CHICKEN-CURRY4	&(cc4)	& 915	& 66 \\ \hline
	TARAMOSALATA-1	&(t1)	& 908	& 61 \\ \hline
	PORK-BRAISE-1	&(pb1)	& 908	& 48 \\ \hline
	CHICKEN-WINE	&(cw)	& 908	& 52 \\ \hline
	CATFISH-BOIL	&(cfb)	& 908	& 55 \\ \hline
	PICADILLO	&(pd)	& 900	& 92 \\ \hline
\end{tabular}
&
\begin{tabular}[t]{|rl||r|r|}\hline
	\multicolumn{2}{|c||}{\bf document}	
		& \multicolumn{1}{|c}{\bf score}	
			& \multicolumn{1}{|c|}{\bf size} \\ \hline\hline
	BOUILLABAISSE	&(bb)	& 893	& 165 \\ \hline
	WATERCRESSSOUP	&(wcs)	& 886	& 88 \\ \hline
	SCALLOPS-1	&(s1)	& 872	& 79 \\ \hline
	CAJUN-LAMB	&(cl)	& 865	& 65 \\ \hline
	MEAT-CURRY	&(mc)	& 858	& 103 \\ \hline
	LEG-OF-LAMB-3	&(ll3)	& 843	& 78 \\ \hline
	CHICKEN-VINDAL	&(cv)	& 843	& 94 \\ \hline
	CHICKEN-KORMA	&(ck)	& 836	& 105 \\ \hline
	WIGILIA-2	&(w2)	& 672	& 57 \\ \hline
	SEVICHE		&(sv)	& 672	& 67 \\ \hline
	SPAGH-SAUCE-2	&(ss2)	& 665	& 48 \\ \hline
	RELISH-1	&(r1)	& 658	& 41 \\ \hline
	FISH-CHOWDER	&(fc)	& 643	& 53 \\ \hline
	EGGPLANT-3	&(e3)	& 643	& 47 \\ \hline
	CHICKEN-YOGURT	&(cy)	& 643	& 86 \\ \hline
	CHICKEN-MOLE-2	&(cm2)	& 643	& 52 \\ \hline
	CHICKEN-MICRN	&(cm)	& 643	& 54 \\ \hline
\end{tabular}
\end{tabular}
\end{center}
\caption{{\bf Results of {\sc waisindex} Search}}
\label{gfro}
\end{table}
\normalsize


\scriptsize
\normalsize


\section{Visualization and browsing}

The final step in automatic document analysis is the interactive presentation and exploration of results.
A subset of facets can be chosen and starting from these,
a local environment of related items can be explored.
As an additional method for handling large amounts of data,
a 3D browser can then be used to navigate in the information space.

\subsection{Interactive Browsing}

When browsing a very large data repository,
it is desirable to select a set of starting objects
and to int3eractively explore their neighborhood.
This process consists of the following steps:

\begin{description}
	\item[Initialization] \mbox{ }
	\begin{enumerate}
		\item The facets 
			$\left\{ \sigma_i \colon M_i \mid 1 \leq i \leq n \right\}$
			are evaluated with respect to the data acquired in phase one
			and transformed
			into a binary relations (formal contexts)
			$\kappa_i = \langle G,M_i,I_i \rangle, 1 \leq i \leq n$.
		\item The evaluated facets 
			$\kappa_i \mid 1 \leq i \leq n$
			are composed into a single 
			binary relation
			$\kappa = \langle G,M,I \rangle, M = M_1 + \cdots + M_n$
			using the operation of {\em apposition\/}
			(this operation requires the contexts to share a common object set).
		\item A global analysis is performed on the total context $\kappa$,
			chiefly in terms of the collection of local neighborhood concept lattices.
	\end{enumerate}
	\item[Browse Loop] \mbox{ }
	\begin{enumerate}
		\item The local neighborhood of a given seed object is analyzed and previewed.
			To simplify the visualization data to be presented to the user
			(and possibly reach an acceptable number of concepts),
			the local neighborhood is modified using various means:
			raising the connectivity threshold,
			rank-ordering the attributes and restricting to the most important ones,
			restricting to a ball around the seed induced by a similarity metric,
			etc.
		\item The local neighborhood is visualized.
			At this time the user may want to visualize
			the union context of the local neighborhoods
			for the old seed and the new seed
			--- this allows comparison of ``distance'' moved
			and things in common.
		\item Finally,
			a new seed is chosen.
			This may be either an object or an attribute.
	\end{enumerate}
\end{description}

\subsection{The visualization process}

Currently, the interactive browsing component of {\sc wave}
uses a conventional  graph layout algorithm
to compute coordinates of information nodes
on a plane, which are rendered as three-dimensional
objects. Size, shape and color of these objects can be varied
to reflect semantic like e.g.\ relevance or document size. 
Being able to freely navigate in the 3-dimensional space 
provides an intuitive fisheye-like
means of centering in on interesting sub-areas of the graph.
Future work will be directed at true three-dimensional layout
of information like e.g.\ the ``information cones trees'' \cite{RCM93}.


\clearpage

	\section{Conclusions}




	\clearpage
	\tableofcontents
\end{document}